\documentstyle[preprint,aps]{revtex}
\begin{document}
\draft
\title{\bf STRINGY BLACK HOLES AND ENERGY CONDITIONS}
\author{Sayan Kar \thanks{Electronic Address :
sayan@.iopb.ernet.in} \footnote{Present Address : IUCAA,
Post Bag 4, Ganeshkhind, Pune 411007, INDIA\\ Electronic
Address : sayan@iucaa.ernet.in}}
\address{Institute of Physics\\
Sachivalaya Marg, Bhubaneswar--751005, INDIA}
\maketitle
\parshape=1 0.75in 5.5in
\begin{abstract}
The energy condition inequalities for the matter stress energy comprised
out of the dilaton and Maxwell fields in the dilaton--Maxwell gravity theories
emerging out of string theory are examined in detail. In the simplistic
$1+1$ dimensional models, $R\le 0$ (where $R$ is the Ricci scalar), 
turns out to be the requirement for ensuring focusing of timelike geodesics.
In $3+1$ dimensions,
we outline the requirements on matter for pure dilaton theories--these
in turn constrain the functional forms of the dilaton. Furthermore, in
charged dilaton gravity a curious opposite behaviour of the
matter stress energy w.r.t the violation/conservation of the
Weak Energy Condition is noted for 
the electric and magnetic black hole metrics written in the 
string frame of reference.  We also investigate the
matter that is necessary for creating certain specific non--asymptotically
flat black holes. For the electric and magnetic black hole metrics, strangely,
 matter 
satisfies the 
Weak Energy condition in the string frame. 
Finally, the Averaged
Null Energy Condition is evaluated along radial null geodesics for
each of these black hole spacetimes. 
\end{abstract}


\newpage

\section{Introduction}

The low energy effective theory that emerges out of full string theory
by imposing quantum conformal invariance in the world-sheet sigma model
and thereby equating the one--loop beta functions for the metric
and matter couplings to zero, largely
resembles General Relativity (GR) with some new `matter' fields such
as the dilaton, axion etc. {\cite{gsw:book}}.
The field equations (which are the
ones obtained by equating the one--loop beta functions to zero) can therefore
be solved with different ansatzen for the metrics and matter fields. These
solutions thus represent allowed backgrounds for string propagation
--the 'allowance' being the fact that quantum conformal invariance
is satisfied on the worldsheet. Presently, there do exist many solutions
of these
equations representing black--holes {\cite{sbh:ref}}, cosmologies 
{\cite{sc:ref1},\cite{sc:ref2},\cite{sc:ref}} etc.

An important fact about the low--energy theory is that there exists
two different frames in which the features of the spacetime
may look very different. These frames which are known as the 
'Einstein frame' and the 'string frame' are related to each
other by a conformal transformation 
($g_{\mu\nu}^{E} = e^{-2\phi}g_{\mu\nu}^{S}$)
 which involves the massless
dilaton field as the conformal factor. The existence of two different
frames is, however a known feature in certain modifications
to Einstein's theory. Infact the 'string frame' is actually similar
to the Brans--Dicke frame in the well--known Jordan--Brans--Dicke
theory. In the context of string theory one says that the string
`sees' the string metric (which is the metric written in the string frame).
Several of the important symmetries of string theory also rely on the
choice of the string frame or the Einstein frame. For instance, the
familiar $T$ duality {\cite{td:ref}} transformation relates metrics
in the string frame only, whereas $S$
duality {\cite{sd:ref}}
is a valid symmetry only if the equations are written in the
Einstein frame. 

It has been mentioned several times in the literature {\cite{hs:ictp92}},
{\cite{h:ictp92}} that the 
metric in the string--frame violates the inequality $R_{\mu\nu}{\xi}^{\mu}
{\xi}^{\nu} \ge  0$ and hence also the assumption of a local Energy
Condition. Therefore, it has been argued, that the Singularity
Theorems of GR {\cite{wald:gr}} {\cite{he:gr}}
are not valid for the low--energy theory emerging
out of string theory. This is because the Singularity theorems assume
an Energy condition and such an assumption essentially leads to
the concept of geodesic focussing -- a conclusion resulting out of
an analysis of the Raychaudhuri equation {\cite{akr:prd55}}, {\cite{tip:prd77}}. 
The focussing theorem alongwith some additional assumptions
essentially imply the existence of spacetime singularities.

In this paper, we explicitly examine several black hole
geometries in two and four dimensions with regard to the
Weak Energy Condition (WEC)
 $T_{\mu\nu}{\xi}^{\mu}{\xi}^{\nu} \ge 0$. We shall 
point out the domains of violation of these
Energy Conditions and also attempt to arrive at some general statements.
It will turn out that there are black holes which require 
a violation of these conditions as well as solutions which
do not violate them. Moreover, we shall demonstrate 
solutions which satisfy the Weak Energy Condition (WEC) 
and also check the Averaged Null Energy Condition (ANEC)
for several black holes. It will turn out that there are 
quite a few geometries for which the ANEC integral along radial null
geodesics is positive definite. With all these we will try and conclude
that it is somewhat premature to make statements about 
the non--validity of the singularity theorems for the
class of theories emerging out of full string theory.

The paper is organised as follows. In Sec II we analyse the
Energy conditions and geodesic focussing for $1+1$ dimensional
theories of gravity. Sec III deals with the conditions for
Einstein--dilaton gravity. Charged dilaton black holes are
discussed in the fourth section. The ANEC integral is checked in
Sec V. Finally, Sec VI contains a
summary of the main results of the paper.

The sign conventions followed in this paper are those of
Misner, Thorne and Wheeler {\cite{mtw:free}}.
\section{Dilaton Gravity in $1+1$ Dimensions}

We begin with the simple models of gravity in $1+1$ dimensions
{\cite{h:ictp92},\cite{hs:ictp92}}
where the first stringy black hole was discovered.  
Before we explicitly relate $R_{\mu\nu}{\xi}^{\mu}{\xi}^{\nu}$
with the quantities involving the dilaton (using the one loop
$\beta$ function equations) let us look at the focusing 
conditions emerging out of an analysis of the Raychaudhuri
equation.

The Raychaudhuri equation for timelike geodesic congruences in 
a $1+1$ dimensional spacetime, discussed earlier in {\cite{mann:rhcd}},
turns out to be given as :

\begin{equation}
\frac{d\theta}{d\lambda} + \theta^{2} = -R_{\mu\nu}{\xi}^{\mu}{\xi}^{\nu}
\end{equation}

Note that a way to arrive at this equation without starting from first 
principles(as is done in {\cite{mann:rhcd}}) is to substitute appropriate
values of $N$ (background spacetime dimensions--in this case $N=2$) and
$D$ (dimensions of the embedded geometric object--in this case $D=1$) in the
generalised Raychaudhuri equation for families of $D$ dimensional surfaces
in an $N$ dimensional background (for details see {\cite{cg:prd95}}).

Now recall that in $1+1$ dimensions we have 

\begin{equation}
R_{\mu\nu} = \frac{1}{2}g_{\mu\nu}R
\end{equation}

Therefore, we can rewrite the above equation as follows :

\begin{equation}
\frac{d^2 F}{d\lambda^{2}} + (\frac{1}{2}Rg_{\mu\nu}{\xi}^{\mu}{\xi}^{\nu}
 ) F = 0
\end{equation}

where $\theta = \frac{F^{\prime}}{F}$
If the ${\xi}^{\mu}$ are timelike then we need 

\begin{equation}
R \le 0
\end{equation}

in order to have the existence of zeros in a solution of (3).
Such zeros essentially imply a divergence in $\theta$. Hence, 
a converging ($\theta$ negative)
 timelike geodesic congruence must
necessarily come to a focus ($\theta \rightarrow -\infty$
 within a finite value of the affine 
parameter $\lambda$. 
If, $R=0$,then
of course the R.H.S. of the Raychaudhuri equation is identically
zero and we always get a focusing effect. Note that
for $R= 0$ the Raychaudhuri equation
has solutions $F= constant$ and $F= a\lambda + b$. The former 
results in an expansion which is everywhere zero, while, with the
latter one can arrive at focusing. 

In $1+1$ dimensions null geodesics have a unique behaviour.
It can be shown that $\theta$ is identically equal to zero.
This is largely due to the fact that in $1+1$ dimensions 
all metrics are conformally flat and therefore null geodesics
are the same as that for flat $1+1$ dimensional spacetime.
A discussion on this can be found in {\cite{viss:95}}.  

Let us now go back to timelike geodesics.
The question to ask now is whether the two--dimensional black hole 
metrics known to us satisfy the condition $R\leq 0$. 
To see this we have to investigate 
some cases explicitly. Note, however, that in $1+1$ dimensions 
a Weyl rescaling of the metric leaves 
the 'Brans--Dicke' form  of the action invariant. Therefore, it is
perhaps not worth referring to the `Einstein frame' as such  because
it is essentially defined as a frame in which the theory is in the
canonical form.

In pure dilaton gravity with a cosmological constant we can show 
from the field equations that

\begin{equation}
R = 4\left \{ \lambda^{2} - \left ( {\nabla \phi}\right )^{2}
\right \}
\end{equation}

Therefore, if $\lambda$ is zero $R$ is always positive/negative depending 
on what $\phi$ is functionally.
This implies that the focussing may or may not take place in the solutions 
of such a theory. 

Let us now concentrate on some examples of black holes in
two dimensional dilaton gravity.

The two--dimensional black hole metric of Mandal, Sengupta and Wadia
{\cite{sbh:ref}} is given by the metric

\begin{equation}
ds^{2} = - (1 - ae^{Q r}) dt^{2} + \frac{1}{(1 - ae^{Q r})}dr^{2}
\end{equation}

This is a solution in pure  dilaton gravity in $1+1$ dimensions
(i.e. without a cosmological constant ).
The Ricci scalar for this geometry turns out to be :

\begin{equation}
R =  aQ^{2}e^{Qr}
\end{equation}
which is clearly positive. Therefore, even though there is a singularity
at $r=\infty$ focusing within a finite value of the affine parameter
for an initially converging timelike geodesic congruence does not occur.

We now turn to the {\em exact} metric (exact in the sense of full
string theory) due to Dijkgraaf,Verlinde, Verlinde {\cite{dvv:npb}}
(see also {\cite{pt:prl}}) is given as :

\begin{equation}
ds^{2} = 2(k -2)\left [ -\beta (r) dt^{2} + dr^{2} \right ]
\end{equation}

where $\beta (r) = \left ( \coth^{2} r - \frac{2}{k} \right )^{-1}$
and the dilaton is given as :

\begin{equation}
\phi = \phi_{0} + \frac{1}{2} \ln \vert \sinh^{2} \frac{2r}{\beta} \vert
\end{equation}

The quantity $k$ stands for the Kac--Moody level. It is related to the
central charge $c$ of the Wess--Zumino--Witten model based on the group
$SO(2,1)$ gauged by the subgroup $SO(1,1)$ (which is an exact conformal
field theory description of the Witten solution quoted below) by the
relation $c = \frac{3k}{k-2} - 1$ . $k$ takes the value $\frac{9}{4}$
for a bosonic string background for which  $c = 26$.

The Ricci scalar turns out to be :

\begin{equation}
R = \frac{1}{2(k-2)} \frac{2 cosech^{2} r}{\left ( \coth^{2} r - \frac{2}{k}
\right )^{2}} \left [ \frac{2}{k} + 2\coth^{2} r \left (1 - \frac{3}{k}\right )
\right ]
\end{equation}

The positivity/negativity of $R$ crucially depends  
on the value of $k$. If $k \ge 3$ then $R$ is always 
positive. For $k$ lying between $2$ and $3$ there is
always a domain in which $R$ is negative as is easily
noticeable from the expression.

In the $k \rightarrow \infty$ limit the metric goes over to
the Witten black hole (modulo the overall factor $k-2$).
This is given by the metric 

\begin{equation}
ds^{2} = -\tanh^{2} r dt^{2} +  dr^{2}
\end{equation}

which is related by a coordinate transformation to the 
metric discussed first in this sequence (the Mandal, Sengupta,
Wadia black hole).

The Ricci scalar is given as :

\begin{equation}
R = \frac{4}{\cosh^{2} r}
\end{equation}

which, once again is positive. However, note that the metric written
in the above form does not have a singularity anywhere. The point $r=0$
is actually a coordinate singularity. One can, following Witten, do
a Kruskal extension of the geometry and arrive at the form :

\begin{equation}
ds^{2} = -\frac{du dv}{1 - uv}
\end{equation}

where $2u=-{e^{r^{\prime} - t}}$, $2v=e^{r^{\prime} + t}$ and 
$r^{\prime} = r + \ln \left (1-e^{-2r}\right )$.
The denominator in the Ricci scalar now gets replaced by $1-uv$ and 
$uv = 1$ corresponds to the divergence of $R$ and therefore is a
real singularity. However even if we consider the maximally extended
metric the Ricci scalar is still positive definite and timelike
geodesic congruences do not focus.
This is in accordance with the discussion regarding focusing 
on the Mandal, Sengupta, Wadia form of this solution presented
earlier in this section.

There do exist a multitude of other solutions in $1+1$ dimensions
such as the ones
derived in {\cite{np:npb}} and
{\cite{mc:npb}} 
for which one can do a similar analysis. Since the basic idea is the
same we refrain from repeating the same exercise here.

\section{Pure Dilaton Theories in $3+1$ dimensions}

\subsection{Energy conditions}

Before we embark on an analysis of the matter sector of dilaton gravity
theories let us briefly recall the content of the various Energy Conditions
which we shall be checking out for specific solutions.
Below, $T_{\mu\nu}$ represents the energy momentum tensor for matter
and $\xi^{\mu}$ represents a timelike or null vector as specified
in the respective conditions.

(1) {\bf Local energy conditions}

{\em (a) Strong Energy Condition}

\begin{equation}
\left ( T_{\mu\nu} - \frac{1}{2}g_{\mu\nu}T \right ){\xi}^{\mu}{\xi}^{\nu}
\ge 0 \qquad \forall \quad timelike \quad {\xi}^{\mu}
\end{equation}

{\em (b) Weak Energy Condition}

\begin{equation}
\left ( T_{\mu\nu}{\xi}^{\mu}{\xi}^{\nu} \right ) \ge 0 \qquad \forall 
\quad timelike \quad {\xi}^{\nu}
\end{equation}

{\em (c) Null Energy Condition}

\begin{equation}
\left ( T_{\mu\nu} k^{\mu}k^{\nu} \right ) \ge 0 \qquad \forall \quad null
\quad k^{\mu}
\end{equation}

(2) {\bf Global Energy conditions} {\cite{tip:prd77}}

{\em  (a) Averaged Weak energy Condition}

\begin{equation}
\int_{-\infty}^{\infty} T_{\mu\nu}{\xi}^{\mu}{\xi}^{\nu} d\lambda \ge 0
\end{equation}

where the integration is over a complete, timelike geodesic in the spacetime.

{\em (b) Averaged Null Energy condition}

\begin{equation}
\int_{\infty}^{\infty} T_{\mu\nu} k^{\mu}k^{\nu} d\lambda \ge 0
\end{equation}

where the integration is now over a complete, null geodesic in the spacetime.

\subsection{Checking the Energy Conditions for Dilaton Gravity}

We now check the various Energy conditions listed above for the theory
of dilaton gravity in $3+1$ dimensions.

The action integral for the Einstein--dilaton--Maxwell theory is given
as :

\begin{equation}
S_{EDM} = \int d^{4}x \sqrt{-g} e^{-2\phi}\left [R + 4g_{\mu\nu}\nabla^{\mu}
\phi \nabla^{\nu} \phi - \frac{1}{2}g^{\mu\lambda}g^{\nu\rho}
F_{\mu\nu}F_{\lambda
\rho} \right ]
\end{equation}

Varying with respect to the metric, dilaton and Maxwell fields we get the 
field equations for the theory  given as:

\begin{eqnarray}
R_{\mu\nu} = -2\nabla_{\mu}\nabla_{\nu}\phi + 2F_{\mu\lambda}F_{\nu}^{\lambda}\\
\nabla^{\nu}\left ( e^{-2\phi}F_{\mu\nu} \right ) = 0 \\
4\nabla^{2}\phi - 4 \left ( \nabla \phi \right )^{2} + R - F^{2} = 0
\end{eqnarray}

These equations are also the $\beta $ function equations for
a worldsheet sigma model obtained by imposing quantum conformal
invariance and setting the $\beta$ functions to zero.
Note that without the Maxwell field we have essentially
a Brans--Dicke type theory with the Brans--Dicke parameter
explicitly set to $\omega = -1$.

Consider the very first equation (without the Maxwell field)
and recast it in the form of the
Einstein equation $G_{\mu\nu} = e^{2\phi}T_{\mu\nu}$. This is
the generic form of the Einstein equation for a Brans--Dicke
type theory (for details see Weinberg{\cite{wein:71}}
). The $T_{\mu\nu}$ contains a part $T_{\mu\nu}^{\phi}
$ and a part $T_{\mu\nu}^{M}$ where $M$ stands for all matter
apart from the dilaton.
With this, one can now
write down the Energy momentum tensor for the dilaton field which turns 
out to be

\begin{equation}
T_{\mu\nu}^{\phi} = e^{-2\phi}\left [
-2\nabla_{\mu}\nabla_{\nu} \phi + g_{\mu\nu}\nabla^{2}\phi \right ]
\end{equation}

Therefore, the various Energy conditions turn out to be equivalent to the
following inequalities :

{\bf WEC :} 

\begin{equation} 
T_{\mu\nu}^{\phi}{\xi}^{\mu}{\xi}^{\nu} = -e^{-2\phi}
\left [ 2{\xi}^{\mu}{\xi}^{\nu}\nabla_{\mu}
\nabla_{\nu} \phi + \nabla^{2}\phi \right ] \ge 0
\end{equation}

For $\xi^{\mu}$ along a geodesic curve the first term inside the
square brackets can be shown to reduce to
$2\frac{d^{2}\phi}{ds^{2}}$. Since from the field equations we have
$\nabla^{2}\phi = 2\left (\nabla \phi \right ) ^{2} $ the WEC reduces to
$2\phi^{\prime\prime} +  2\left (\nabla \phi \right ) ^{2} \le 0 $

{\bf AWEC :}

\begin{equation}
\int^{\infty}_{-\infty} e^{-2\phi}\left ( 2\phi^{\prime\prime} + 2
 \left (\nabla \phi \right )^{2} \right ) d\lambda \le 0
\end{equation}

{\bf NEC :}

\begin{equation}
T_{\mu\nu}^{\phi}{\xi}^{\mu}{\xi}^{\nu} = -2 e^{-2\phi} k^{\mu}k^{\nu}
\nabla_{\mu}\nabla_{\nu}\phi
\end{equation}

For $k^{\mu}$ a tangent along a null geodesic we have 
the requirement :

\begin{equation}
\phi^{\prime\prime} \le 0
\end{equation}

{\bf ANEC :}

\begin{equation}
\int_{-\infty}^{\infty}T_{\mu\nu}^{\phi}k^{\mu}k^{\nu}d\lambda =
2\int_{-\infty}^{\infty} e^{-2\phi}\phi^{\prime\prime}d\lambda \le 0
\end{equation}

These are the constraints which the dilaton field has to obey in order to
satisfy the respective Energy conditions.

In order to understand the constraints on $\phi$ better we write down the
WEC inequalities explicitly for the general, static spherisymmetric metric
given by :

\begin{equation}
ds^{2} = -e^{2\psi (r)}dt^{2} + \frac{dr^{2}}{1-\frac{b(r)}{r}}
+ r^{2}\left (d\theta^{2} + \sin^{2}\theta d\phi^{2}\right )
\end{equation}

where $b(r)$ and $\psi(r)$ are two unknown functions.

The $\rho \ge 0$, $\rho +\tau \ge 0$ and $\rho +p\ge 0$ inequalities 
turn out to be :

\begin{eqnarray}
-2e^{-2\phi}\left ( 1-\frac{b(r)}{r}\right ) \left [ {\phi^{\prime}}^{2} -
\phi^{\prime}\psi^{\prime} \right ] \ge 0 \\
-4e^{-2\phi}\left ( 1-\frac{b(r)}{r}\right ) \left [ {\phi^{\prime}}^{2} -
{\phi^{\prime}\over r} \right ] \ge 0 \\
2e^{-2\phi}\left ( 1 -\frac{b(r)}{r}\right ) \phi^{\prime}\left [ \psi^{\prime}
-\frac{1}{r} \right ] \ge 0
\end{eqnarray}
 
The above inequalities are obtained using the expression for $T_{\mu\nu}^{\phi}$
in terms of the scalar field and its derivatives (Eqn (22)). 
These are the constraints on $\phi$, $\psi$ and $b$ which must be
obeyed if we believe in the WEC. We now analyse a couple of choices
of $\phi$.

If $\phi$ is linear in $r$ i.e. $\phi(r) = r$ say, then we find that the
requirements turn out to be $\psi^{\prime} \ge \frac{1}{r}$, $ \frac{1}{r}
\ge 1$ and $\psi^{\prime}\ge 1$.
Clearly as $r\rightarrow \infty$ the second condition is violated. Thus with
a linear dilaton field one always will end up with a violation as
$r\rightarrow \infty$. In contrast, if $\phi = \ln r$ then the first and third 
inequalities reduce to the requirement $\psi^{\prime}\ge \frac{1}{r}$.
whereas the second one is identically satisfied (L.H.S. of the second
WEC inequality is zero. Thus with a logarithmic dilaton --which means a 
linear string coupling-- the WEC can be satisfied everywhere!

\section{ Energy Conditions for Electric and Magnetic Black Holes
in Dilaton--Maxwell Gravity}

In this Section we focus our attention on the Energy Condition inequalities
for the electric and magnetic black--hole solutions in dilaton--Maxwell
gravity {\cite{gm:npb88}}, {\cite{ghs:prd91}}.

Evaluating the Einstein tensor $G_{\mu\nu}$ for the general, static, 
spherisymmetric
metric quoted in the previous section
we can write down the expression for the
Energy Conditions. This is equivalent to looking at the matter
energy momentum tensor because we are dealing with exact solutions
of the field equations.

For a diagonal $T_{\mu\nu} \equiv \left (\rho (r), \tau (r), p(r), p(r)\right )$
we find that the WEC reduces to the following inequalities :

\begin{equation}
\rho \ge 0 \quad ; \quad \rho +\tau \ge 0 \quad ; \quad \rho + p \ge 0
\end{equation}

Note here that the NEC consists of only the second and third inequalities. 
From the Einstein equations we therefore end up with the following
conditions on $b(r)$, $\psi(r)$  and its derivatives.

\begin{equation}
\rho (r) = \frac{b^{\prime}}{r^{2}} \ge 0 
\end{equation}
\begin{equation}
\rho (r) + \tau (r) = \frac{b^{\prime}r - b}{r^{3}} + \frac{2\psi^{\prime}}{r}
\left ( 1-\frac{b(r)}{r} \right ) \ge 0 
\end{equation}
\begin{equation}
\rho (r) + p(r) = \frac{b + b^{\prime}r }{2r^{3}} + \left ( 1-\frac{b(r)}{r}
\right ) \left [ \psi^{\prime\prime} + {\psi^{\prime}}^{2} +
\frac{\psi^{\prime}}{r} + \frac{b- b^{\prime}r}{2r(r-b)}\psi^{\prime} \right ] \ge 0 
\end{equation} 

We have absorbed the factor $e^{2\phi}$ in a redefinition of the
components of $T_{\mu\nu}$ (i.e.$ \rho = e^{2\phi} \bar \rho$ and so on
-- where $\bar \rho $ is the actual component of the energy--momentum
tensor).
This factor will however have to be brought
back when we discuss the averaged versions of the Energy Conditions.
For a discussion of the local conditions the overall $e^{2\phi}$ is
irrelevant.

We will now choose the explicit functional forms of $b(r)$ and $\psi (r)$ which
correspond to the well--known black hole solutions in
string theory and thereby
check out the WEC inequalities for each of them. 

\subsection{Electric Black hole}

The metric (in the string frame)
 and matter fields which solve the dilaton--Maxwell--Einstein
field equations Eqns. 19--21) to yield the electric black hole are given as :

\begin{equation}
ds^{2} = -\left (1-\frac{2m}{\hat r}\right ) \left ( 1+\frac{2m\sinh^{2}\alpha
}{\hat r}\right )^{-2} dt^{2} + \frac{d{\hat r}^{2}}{1-\frac{2m}{r}}
+ {\hat r}^{2}d\Omega^{2}
\end{equation}

\begin{eqnarray}
A_{t} = -\frac{m\sinh 2\alpha}{\sqrt 2 \left [ \hat r + 2m\sinh^{2}\alpha
\right ] } \\
e^{-2\phi} = 1+ \frac{2m\sinh^{2}\alpha}{\hat r}
\end{eqnarray}

The geometry has a horizon at $\hat r = 2m$ and a singularity at $r=0$.
Identifying the functions $\psi$ and $b$ with the  metric coefficients in
the above expression we can now write down the WEC inequalities for this
black--hole geometry.

Firstly, note that $\rho = 0$ because $b(r)=2m$ which is a constant.
The other two inequalities turn out to be :

\begin{equation}
\rho + \tau = \frac{4m}{{\hat r}^{3}} \frac{(1-\frac{2m}{\hat r})
\sinh^{2}\alpha}{1+\frac{2m}{\hat r}\sinh^{2}\alpha} \ge 0
\end{equation}

\begin{eqnarray}
\rho + p = \frac{1}{m^{2}} \frac{2x^{3}\sinh^{2}\alpha}
{(1+2x\sinh^{2}\alpha)^{2}} \nonumber \\ 
\left [ (5x - 1) + 
\sinh^{2}\alpha 2(x (1+x) ) \right ] \ge 0
\end{eqnarray}

where in the last equation $x=\frac{m}{\hat r}$.
We now have to check whether these inequalities are satisfied or violated.
It is easy to comment on the first one -- for all $\hat r \ge 2m$ it is
satisfied. The nature of the second inequality can be
understood as follows. 

Firstly, the prefactor outside the square brackets is positive. Therefore,
the sign of the full expression depends entirely on the sign of the
term in square brackets. Note that this is a quadratic form in $x$. It can be
written as :

\begin{equation}
(x-x_{1})(x-x_{2}) 
\end{equation}

where $x_{1},x_{2}$ are the two roots of the function equated to zero.
The explicit forms of the roots are :

\begin{equation}
x_{1,2} = \frac{-2\sinh^{2}\alpha - 5 \pm \sqrt{\left ( 2\sinh^{2}{\alpha}
+ 5\right ) ^{2} + 8\sinh^{2}{\alpha}}}{4\sinh^{2}{\alpha}}
\end{equation}

Therefore, $x_{1}$ (+sign) is always positive while $x_{2}$ is entirely
negative. Hence in order to have $(x-x_{1})(x-x_{2}) \ge 0$ we require

\begin{equation}
x\ge x_{1} =  \frac{-2\sinh^{2}\alpha - 5 + \sqrt{\left ( 2\sinh^{2}{\alpha}
+ 5\right ) ^{2} + 8\sinh^{2}{\alpha}}}{4\sinh^{2}{\alpha}}
\end{equation}

We also note the following :

(i) Violation occurs in the region of small $x$ (i.e. large $\hat r$).
However as one approaches $\hat r \rightarrow \infty$ we find that
the amount of violation becomes smaller. Infact, between a certain 
$x = x_{0}$ and $x = 0$ there is a point where the L. H. S. of the
full WEC inequality
(including the prefactor) has a
minimum. This indicates the maximum negative value it can take.

(ii) For $\alpha$ increasing we note that the domain over which
violation occurs (in $x$) becomes smaller --therefore for $\alpha$
very large it can actually become miniscule in $x$ (and therefore
very large in $\hat r$.

(iii) The string coupling which is given by $e^{\phi}$ is of the form:

\begin{equation}
e^{\phi} = \frac{1}{\sqrt{1+2x\sinh^{2}\alpha}}
\end{equation}

This is a monotonically decreasing function of $x$. The string
coupling is large where WEC violation occurs and it is small in the
region where the WEC is satisfied. One cannot however conclude from
this that the strength of the coupling is a sort of measure for 
WEC violation/satisfaction.

We now turn to an analysis of the WEC inequalities for the extremal
limit of the electric black hole.

The extremal limit is obtained by taking the following limit for the
parameters appearing in the electric black hole.

\begin{equation}
\quad m\rightarrow 0 \quad , \quad \alpha \rightarrow \quad \infty
but \quad m\cosh^{2}\alpha \quad fixed
\end{equation}

The line--element turns out to be :

\begin{equation}
ds^{2} = - \left (1+\frac{2M}{\hat r}\right )^{-2} dt^{2} +
d{\hat r}^{2} + {\hat r}^{2} d\Omega^{2}
\end{equation}

with $m\cosh^{2}\alpha = M$.

The WEC inequalities (only $\rho + \tau \ge 0$, $\rho + p \ge 0$
because $\rho = 0$) turn out to be equivalent to :

\begin{equation}
\rho + \tau = \frac{4M}{{\hat r}^{3}}\frac{1}{1+\frac{2M}{\hat r}} \ge 0
\end{equation}

\begin{equation}
\rho + p = -\frac{2M}{{\hat r}^{3}}\frac{1-\frac{2M}{\hat r}}{\left (1+\frac{2M}{\hat r}
\right )^{2}}
\end{equation}

Notice that the third inequality is violated everywhere
(i.e. $\forall \hat r > 2M$) whereas the second one 
is satisfied everywhere.
The lower bound of the domain of WEC violation has now shifted from
$r= r_{0} > 2M$ to $r = 2M$.

\subsection{Magnetic Black Hole} 

In the string frame the dual solution known as the magnetic black hole
is obtained by multiplying the electric metric in the Einstein frame 
by a factor $e^{-2\phi}$ (Note the sign of $\phi$). In a more generalised 
sense this is the S--duality transformation which changes $\phi \rightarrow
-\phi $ and thereby inverts the strength of the string coupling
. Also recall that the magnetic and electric solutions are the same
if looked at from the Einstein frame).
 
Therefore, the magnetic black hole metric is given by :

\begin{equation}
ds^{2} = - \frac{1-\frac{2M}{r}}{1-\frac{Q^{2}}{Mr}}dt^{2} + 
\frac{dr^{2}}{\left ( 1 - \frac{2M}{r}\right ) \left ( 1 - \frac{Q^{2}}{Mr}
\right )} + r^{2}d\Omega^{2}
\end{equation}

Using the same methods as before we write down the three Energy Condition
inequalities for the metric given above.
These turn out to be :

\begin{equation}
\rho = \frac{2Q^{2}}{r^{4}} \ge 0
\end{equation}

\begin{equation}
\rho + \tau = \frac{2Q^{2}}{Mr^{3}}\left ( \frac{2M}{r} - 1 \right ) \ge 0
\end{equation}

\begin{equation}
\rho + p = \frac{Q^{2}}{2Mr^{3}}\left ( \frac{1 -\frac{2M}{r}}{1 - 
\frac{Q^{2}}{Mr}} + 1 \right )  \ge 0
\end{equation}

It is easily seen that the second inequality is now violated for all
$r> 2M$ whereas the third is satisfied for all $r> 2M$. Note that
this is exactly opposite to what happened to the Energy Conditions 
for the electric black hole ! This opposite behaviour is shown in  
Table I given below.

However, one cannot actually make a general statement about the relation
between electric--magnetic solutions (which are dual to each other),
the violation of the Energy Condition inequalities and the strength of the 
string coupling.

Finally, before we move on to other black--hole geometries let us
look at the extremal limit of the magnetic black hole solution.

The metric for the solution in the extremal limit is given as :

\begin{equation}
ds^{2} = -dt^{2} + \frac{dr^{2}}{\left ( 1 - \frac{2M}{r} \right )^{2}}
+ r^{2}d\Omega^{2}
\end{equation}

The energy condition inequalities turn out to be :

\begin{equation}
\rho = \frac{4M^{2}}{r^{4}} \ge 0
\end{equation}

\begin{equation}
\rho + \tau = \frac{4M}{r^{3}} \left ( \frac{2M}{r} - 1 \right ) \ge 0
\end{equation}

\begin{equation}
\rho + p = \frac{2M}{r^{3}} \ge 0
\end{equation}

Therefore, the first and the third inequalities are satisfied for all
values of $r$ whereas the second one is violated only if $r> 2M$.
The geometry has spacelike slices resembling an infinite horn extending
from infinity to $r=2M$. There is no singularity here.

\subsection{Other Black Hole Metrics}

We now move on towards analysing the energy condition inequalities for
certain recently derived non--asymptotically flat black hole solutions in 
dilaton--Maxwell gravity due to Chan, Mann and Horne {\cite{cmh:npb95}}.
 Here we have a surprise in store for us.
We will see that for the electric solution the energy condition
inequalities are satisfied in both the string as well as the Einstein
frame of reference. 

Let us first look at the solution in the Einstein frame. The metric
is given as :

\begin{equation}
ds^{2} = -\frac{1}{\gamma^{4}} \left ( r^{2} - 4\gamma^{2}M \right ) dt^{2}
+ \frac{4r^{2}}{r^{2} - 4\gamma^{2}M}dr^{2} + r^{2}d\Omega^{2}
\end{equation}

with the dilaton and Maxwell fields as 

\begin{equation}
\phi (r) = -\frac{1}{2} \ln 2Q^{2} + \ln r
\end{equation}
 
\begin{equation}
F_{tr} = \frac{Qe^{2\phi}}{\gamma^{2}r}
\end{equation}

Note that the dilaton rolls from $-\infty$ to $+\infty$ as  $r$ changes its
value from $0$ to $\infty$.

As in the previous cases we first write down the string metric by performing
the usual conformal transformation on the metric. This turns out to be

\begin{equation}
ds^{2} = -\frac{r^{2}}{\gamma^{4}} \left ( 1 - \frac{2{\sqrt 2}\gamma^{2}M}
{Qr}\right ) dt^{2} + \left ( 1 -  \frac{2{\sqrt 2}\gamma^{2}M}
{Qr}\right )^{-1} dr^{2} + r^{2}d\Omega^{2}
\end{equation}

For the Einstein metric one can check that the energy conditions are satisfied.
What about the inequalities for the string metric ? Note that since $b(r)$ is a 
constant we have $\rho = 0$ straightaway. 
The other two inequalities turn out to be as follows.

\begin{equation}
\rho + \tau = \frac{2}{r^{2}}\left ( 1 -\frac{A}{r}\right ) \ge 0
\end{equation}

\begin{equation}
\rho + p =  \frac{A}{2r^{3}} + \frac{1}{r^{2}} \ge 0
\end{equation}

where $A = \frac{2\sqrt 2 \gamma^{2} M}{Q}$.

Surprisingly, for all $r\ge A$ all three inequalities are satisfied.
The string coupling $e^{\phi}$ changes from $0$ to $\infty$ as one varies
$r$ from $0$ to $\infty$. 

In a similar way let us look at the string metric for the magnetic black
hole which is given as :

\begin{equation}
ds^{2} = -\frac{2Q^{2}}{\gamma^{4}} \left ( 1 - \frac{4M}{r} \right ) dt^{2}
+ \frac{2Q^{2}}{r^{2}} \left ( 1 - \frac{4M}{r} \right )^{-1} dr^{2}
+ 2Q^{2}d\Omega^{2}
\end{equation}

Since the coefficient of $d\Omega^{2}$ is a constant  here we
cannot  straightaway use the formulae for the WEC in terms of the $b(r)$ and $\psi(r)$.
After some simple algebra we find that the WEC inequalities reduce to :

\begin{equation}
\rho = 
frac{1}{2Q^{2}} \ge 0
\end{equation}

\begin{equation}
\rho + \tau = 0
\end{equation}

\begin{equation}
\rho + p = \frac{\left ( 1-\frac{2M}{r}\right )}{2Q^{2}}
\end{equation}

Note that the first inequality is satisfied for all values of $r$ where
$r=4M$ is the location of the horizon. In contrast, the third inequality
is satisfied for all $r>2M$.
Thus the weak energy condition is satisfied for all
values of $r\ge 2M$. 
These conclusions are shown in a tabular form in Table 2.

\section{The Status of the ANEC for Stringy Black Holes}

For each of the geometries discussed in the previous section we shall
now evaluate the ANEC integral along radial null geodesics.
To do this we need to know the tangent vectors along radial null
geodesics in the general, static, spherisymmetric metric quoted in Sec III.
A choice for $k^{\mu}$ in the coordinate frame is :

\begin{equation}
k^{\mu} \equiv \left (\frac{dt}{d\lambda}, \frac{dr}{d\lambda}, 0, 0
\right ) = \left ( e^{-2\psi}, e^{-\psi}\sqrt{1-\frac{b(r)}{r}}, 0, 0
\right )
\end{equation}

Note that these choices for $\frac{dt}{d\lambda}$ and $\frac{dr}{d\lambda}$
satisfy the geodesic equations and also maintains the null character of the
geodesic.
In the proper frame (we need to go to the proper frame
because the $G_{\mu\nu}$ used earlier in this paper 
are evaluated in this frame--it is entirely
a matter of choice) this tangent vector is transformed to :

\begin{equation}
k^{\hat \mu} \equiv \left (e^{-\psi}, e^{-\psi}, 0, 0 \right )
\end{equation}

where the hat is used to distinguish between the two frames.
 
The ANEC integral therefore becomes :

\begin{equation}
I^{ANEC} = \int_{-\infty}^{\infty} e^{-2\phi}
G_{{\hat \mu}{\hat \nu}}k^{\hat \mu}k^{\hat \nu} d\lambda
= \int_{r_{H}}^{\infty} \left ( \rho + \tau \right ) 
e^{-2\psi-2\phi} \frac{d\lambda}{dr}
dr
\end{equation}
 where $r_{H}$ is the horizon radius.

All we need to do now is to use the expressions for $\rho +\tau$
(implicitly assuming that the $e^{2\phi}$ factor
is removed by a further redefinition) 
and the corresponding functional forms for the
dilaton for each of the black holes discussed previously and evaluate the
integral.

Let us look at the values of the ANEC integrals for the electric and
magnetic asymptoticaly flat black holes. These are :

\begin{eqnarray}
I^{ANEC}_{elec} = \frac{\sinh^{2}\alpha}{m} \left ( \frac{1}{2} + \frac{1}{3}
\sinh^{2}\alpha \right ) \\
I^{ANEC}_{mag} = \frac{Q^{2}}{4M^{3}} \left ( \frac{Q^{2}}{6M^{2}} - 1
\right )
\end{eqnarray}

Notice that the first of these is a positive quantity while
the other one is negative  for all 
$ Q^{2} \le 4M^{2}$. The extremal limits of both these
black holes exhibit the same behaviour as their non--extremal
counterparts.

The values of $I^{ANEC}$ for all these black holes are quoted in
Table III. 

The fact to note here is that apart from the 
magnetic, asymptotically
flat black holes  and their extremal limit all the other
solutions satisfy the ANEC ! 
 
\section{Summary and Outlook}

The above analysis of the WEC inequalities for the well--known
black hole metrics in low energy stringy gravity indicate 
a few essential facts.
We now list them here.

(i) In $1+1$ dimensional dilaton gravity theories the $R\le 0$  condition on 
the Ricci curvature is the analog of the usual energy condition 
which has to be obeyed
in order to ensure focussing of timelike geodesics. 
We have illustrated this with a couple of well
known black hole metrics in $1+1$ dimensional theories which includes
the exact metric due to Dijkgraaf, Verlinde, Verlinde. 

(ii) For $3+1$ dimensional theories with just a dilaton field we have
outlined the conditions on $\phi$ which must be satisfied if matter
has to satisfy an Energy condition. Specific chioces of the dilaton
are used to illustrate the violation/conservation of the WEC. 

(iii) In charged dilaton gravity in $3+1$ dimensions an explicit
evaluation of these inequalities reveal interesting features.
The electric black hole and its magnetic counterpart (in the 
string frame of reference) exhibit quite an opposite behaviour 
as far as violation/conservation of the WEC is concerned. For 
the electric solution violation occurs away from the horizon
and extends upto infinity-- in this region the string coupling
is ofcourse strong. Moreover, it is the $\rho + p$ inequality
which is violated. On the contrary, for the magnetic hole, 
the violation is present near the horizon and occurs only for
the $\rho + \tau$ inequality! These features persist if one
takes the extremal limit in these metrics.

(iv) For certain non--asymptotically flat black hole
solutions recently discovered the electric black holes do not
violate the WEC even in the string metric. Their
magnetic counterparts also satisfy the  Weak Energy Condition!

(v) Several of the stringy black holes seem to satisfy the 
ANEC evaluated along radial null geodesics in the spacetime.
Only the asymptotically flat magnetic black hole and its
extremal limit are the two exceptions.
   
Our aim in this paper has been to analyse in some detail the
nature of the 'matter' that is required to create the well--known
dilaton--Maxwell black hole solutions in low energy effective
string theory. Classical matter which may collapse to form such
black holes must necessarily satisfy the WEC or some other energy
condition. But, if matter violates the Energy Conditions we cannot
conclude that singularities do not exist.
Infact, there are does exist examples of
singular metrics with WEC violating matter {\cite{br:prd}}.

How does one justify the presence of singularites with
the violation of the Energy Conditions within the context of
the Singularity theorems? One can take the attidude that
these theorems are not valid. This is perhaps not
entirely correct because there have been attempts to {\em extend} the
singularity theorems by further{\em weakening/changing}
the assumptions on matter.
A step towards this is the proposal for global Energy conditions.
Singularity theorems with such global conditions have been proved
by Roman {\cite{tr:prd}} and Borde {\cite{bord:cqg87}}.
Some of the geometries discussed in this paper do satisfy the ANEC
along radial null geodesics.
However, for those stringy black holes
which violate the local as well as the global Energy Conditions
one 
should try and extend the singularity theorems with some
assumption on matter which is different from the known ones.

On the other hand, we also seem to have some solutions
which satisfy the WEC without any problems. For such
solutions the original Singularity theorems are obviously valid !

 Therefore, we
might perhaps conclude by saying that it is somewhat
premature to arrive at a general statement on the validity/non--validity of
the singularity theorems in string inspired gravity theories.
Until we can prove a no--go theorem stating that there is 
{\em no} general assumption on matter which can be used to
arrive at the existence of singularities in stringy generalisations
of GR we should leave this as an open issue worth future investigation.


\begin{table}
\caption{ WEC and Electric, Magnetic Solutions }
\begin{tabular}{|c|c|c|c|c|}\hline
{\bf Black Hole} & {\bf WEC2} & {\bf WEC3} & $e^{\phi}\quad$ {\bf at} $\quad r = 0$ & 
$e^{\phi} \quad$ {\bf at} $\quad r\rightarrow \infty$ 
\\ \hline
Electric & Satisfied & Violated $\forall r = r_{0}\ge 2m$ & Weak & Strong \\ 
\hline
Magnetic & Violated  & Satisfied & Strong & Weak \\ \hline
\end{tabular}
\end{table}

\begin{table}
\caption{ WEC and Nonasymptotically flat Electric, Magnetic Solutions }
\begin{tabular}{|c|c|c|c|}\hline
{\bf Black Hole} & {\bf WEC1} & {\bf WEC2} & {\bf WEC3} \\ \hline
Electric & Satisfied & Satisfied $\forall r \ge A $ & Satisfied $\forall r$  
\\ \hline
Magnetic & Satisfied($\forall r$) & Satisfied & Satisfied 
($\forall r \ge 2M$) \\ \hline
\end{tabular}
\end{table}
 
\begin{table}
\caption {Stringy Black Holes and the ANEC}
\begin{tabular}{|c|c|c|}
{\bf Black Hole} & {\bf Value of ANEC Integral} & {\bf Status of ANEC} \\ \hline
Elec., A F & $\frac{\sinh^{2}\alpha}{m}\left ( \frac{1}{2} + \frac{1}{3}
\sinh^{2} \alpha \right ) 
$ & Satisfied \\ \hline
Extr., Elec.& $\infty$ & Satisfied \\ \hline
Mag., AF & $\frac{Q^{2}}{4M^{3}}
\left ( \frac{Q^{2}}{6M^{2}} -1 \right )$ & Violated \\ \hline
Extr., Mag. & $-\frac{1}{3M}$ & Violated \\ \hline
Elec., N--AF & $\frac{Q^{2}\gamma^{2}}{A^{4}}$ & Satisfied \\ \hline
Mag., N--AF & $0$ & Satisfied \\ \hline
\end{tabular}
\end{table} 
 
\end{document}